\documentclass[journal,letterpaper]{IEEEtran}

\usepackage{graphicx,cite,epsfig,amssymb,amsmath,amsfonts,multicol,subfigure,mathtools,bm,mathrsfs,setspace}
\usepackage{multirow}

\newcommand{\tabincell}[2]{\begin{tabular}{@{}#1@{}}#2\end{tabular}}

\begin{document}
\title{\huge Cocktail BPSK: Energy Reused Scheme for High Achievable Data Rates}
\author{
	Bingli~Jiao,
	Yuli~Yang
	and~Mingxi~Yin
	\thanks{B. Jiao ({\em corresponding author}) and M. Yin are with the Department of Electronics and Peking University-Princeton University Joint Laboratory of Advanced Communications Research, Peking University, Beijing 100871, China (email: jiaobl@pku.edu.cn, yinmx@pku.edu.cn).}
	\thanks{Y. Yang is with the Department of Electronic and Electrical Engineering, University of Chester, Chester CH2 4NU, U.K. (e-mail: y.yang@chester.ac.uk).}
}

\maketitle

\begin{abstract}
The present paper proposes a novel transmission strategy, referred to as cocktail BPSK, whereat two independent BPSKs are superposed with the non-orthogonal basis in a parallel transmission. In contrast to the conventional signal superpositions, the proposed scheme avoids the interference between the two symbols, allows the symbol-energy-reuse of each other and gains the extra energy. Based on the formulation of the mutual informations, the theoretical analysis shows that the cocktail BPSK scheme can achieve high data rate beyond the channel capacity at very low SNR, and the numerical results confirm this approach eventually. 

\end{abstract}


\IEEEpeerreviewmaketitle

\section{Introduction}

Along with the development of information theory, the derivation of mutual information plays an important role in theoretical establishment, because it specifies the achievable data rate (ADR) of error-free transmissions through a memoryless additive white Gaussian noise (AWGN) channel by 
\begin{eqnarray}
\begin{array}{l}\label{eq1}
y = x + n,
\end{array}
\end{eqnarray}
where $y$ is the received signal and $x$ is the transmitted signal, and $n$ is the AWGN component from a normally distributed ensemble of power $\sigma_N^2$, denoted by $n \sim \mathcal{N}(0,\sigma_N^2)$.

The ADR of the transmission in (\ref{eq1}) is characterized by the mutual information between $x$ and $y$, expressed as~\cite{Shannon1948}
\begin{equation}\label{eq2}
R = {\rm{I}}(X;Y) = {\rm{H}}(Y) - {\rm{H}}(N),
\end{equation}
where ${\rm{H}}(Y)$ is the entropy of the received signal and ${\rm{H}}(N) = {\log _2} (\sqrt{2 \pi e \sigma_N^2})$ is the entropy of the AWGN.

When the transmitted signals are selected from a Gaussian ensemble of power $\sigma_X^2$, i.e., $x \sim \mathcal{N}(0,\sigma_X^2)$, the channel capacity, i.e., the maximum ADR, reaches
\begin{equation}\label{eqC}
C = \log_2 (1 + \rho) = \log_2 (1 + {\sigma_X^2}/{\sigma_N^2}),
\end{equation}
where $C$ is the channel capacity measured in the unit of [bits/sec/Hz], and $\rho = {\sigma_X^2}/{\sigma_N^2}$ is the signal-to-noise power ratio (SNR) .

\begin{figure}[!t]
	\centering
	\subfigure[]{
		\includegraphics[width=0.4\textwidth]{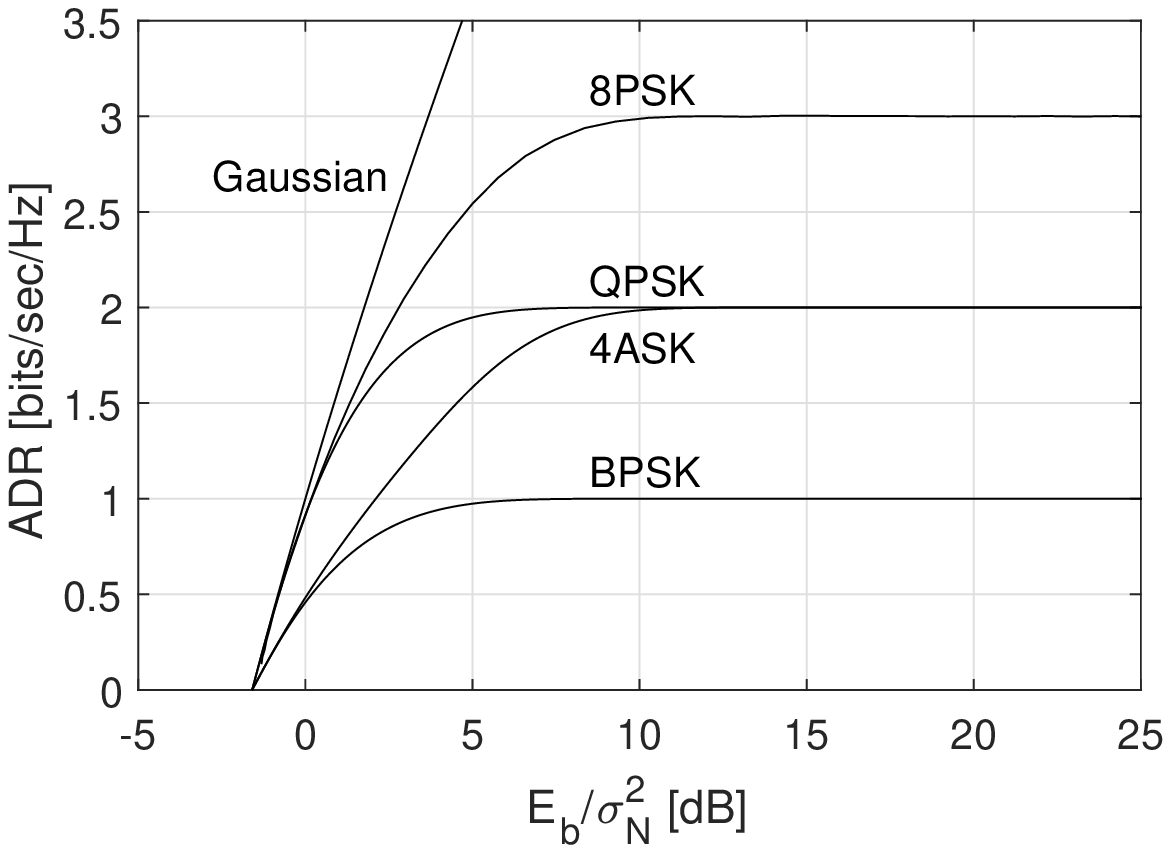}
		\label{fig1a}}
	\subfigure[]{
		\includegraphics[width=0.4\textwidth]{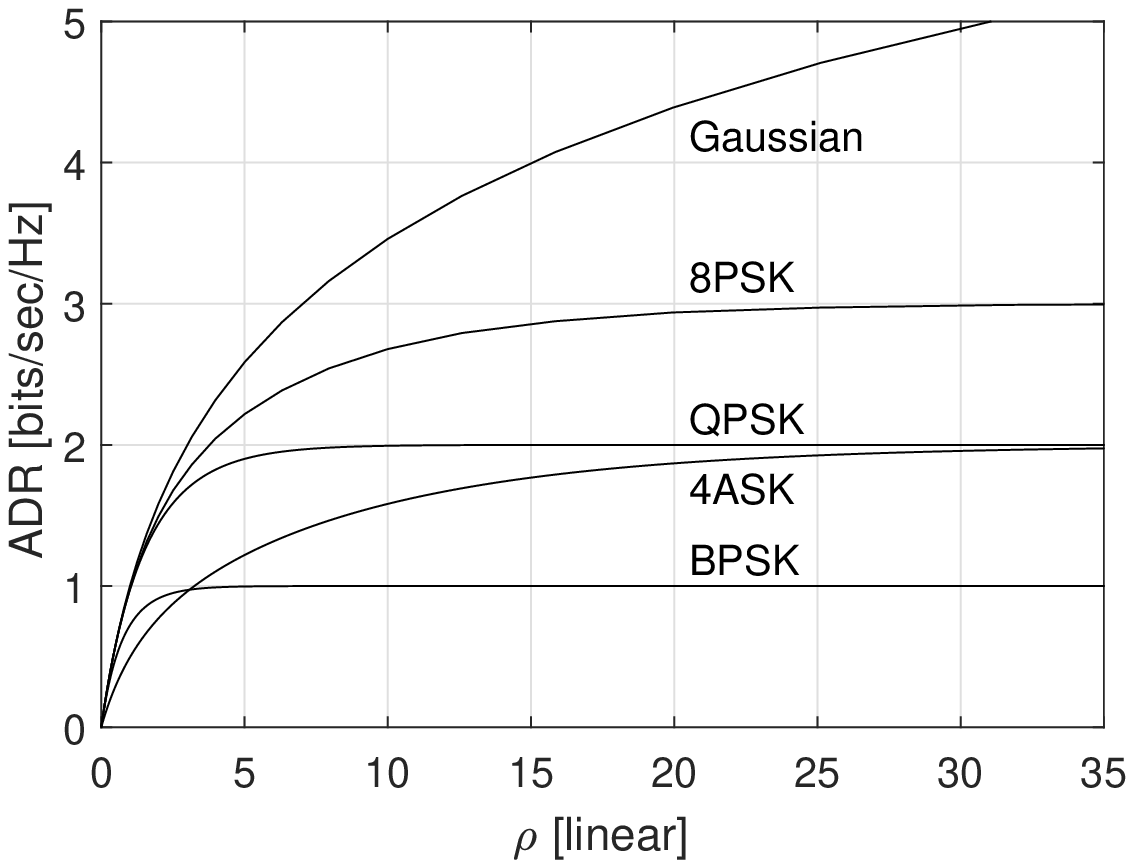}
		\label{fig1b}}
	\caption{Data rates achieved by the Gaussian-distributed input and the modulations of BPSK, 4ASK, QPSK, 8PSK over AWGN channels versus (a) logarithmic ratio of $E_b/\sigma_N^2$ in [dB] and (b) linear ratio of $\rho = \sigma_X^2/\sigma_N^2$.}
	\label{fig1}
\end{figure}

The channel capacity predicts the upper bound of ADRs in concept of error-free transmission. The numerical investigations have been carried out as shown in Fig. {\ref{fig1a}} and (b) for various finite-alphabet inputs using popular modulation schemes, such as ADRs of BPSK, 4ASK, QPSK, and 8PSK. ADRs are plotted versus the logarithmic ratio of energy per bit to that of the noise in decibels ($E_b/{\sigma_N^2}$ [dB]) in Fig. {\ref{fig1a}} and the linear ratio of $\rho$ in (b), where $\rho$ represents the ratio of energy per symbol to that of the noise.  

We remark two features of ADRs at low SNR, i.e., the SNR is in the vicinity of $E_b/{\sigma_N^2} \doteq -1.59$dB {\cite{EbN0}} or $\rho \doteq 0$: i) All the ADRs converge to zero, and ii) derivatives of ADRs of BPSK and QPSK are approximately of the same value ${\log _2}e$ [see Fig. {\ref{fig1b}}]. The second feature comes from the fact that ADRs can be expressed by logarithmic functions of their received SNRs like the channel capacity~\cite{UngerIT, Blahut}. Thus, when the SNR $\rho$ is very low, ADRs are approximately equal to the channel capacity ~\cite{math} 
\begin{equation}\label{eqC2}
C \doteq R \doteq \rho {\log _2}e ,
\end{equation}
with $\rho<<1$, where $C$ is the capacity, $R$ is the ADR of BPSK or QPSK.

For increasing the ADR fundamentally, this work is inspired by the strict concavity of logarithm function that $\log_2(\rho) \leq \log_2(\rho_1) + \log_2(\rho_2)$ with $\rho = \rho_1 + \rho_2$. Concerning this, in~\cite{jiao} we employed multiple data streams in parallel transmissions to convey the information sources for achieving higher ADRs than that of a single data stream. Since both Euclidean geometry and Hamming distance were involved in the formulation, we have not sorted out a specific solution due to the complicity. However, in this work, we track the conception of the parallel transmissions and simplify the formulation in Euclidean geometry only. 

As has been known, splitting a single signal stream into multiple signal steams transmissions has been realized mostly in two manners: One uses orthogonal codes to perform parallel transmissions~\cite{ortho1, ortho2}; however, its sum-data-rate can not be increased at all in comparison with the use of single signal stream, because the orthogonality in Euclidean geometry divides simply the resources of the spectrum and power among signal streams, and each of the split-stream works in the conventional manner eventually. The other method uses the non-orthogonal signal-superposition that provide more degrees of freedom and can enable complicated mathematical operations for improving the transmission rate; however, this method suffers from the inter-stream interference problem, thus, prevents any gain in terms of total ADRs over a single channel.

To the best knowledge of the authors, there has been no method so far that formulates two non-orthogonal streams without any interference and enable the power reuse between them. By creating the method of the signal superposition addressing these two points, we can achieve the high spectral efficiency beyond the channel capacity as explained in the following sections.    

The remainder of this paper is organized as follows. Section II prescribes the basic strategy of the proposed communication scheme and sets up the theoretical framework for the analysis, and section III demonstrates numerical results to confirm this approach. Finally, the paper is concluded in Section IV.

\section{Cocktail BPSK}

This section explains mechanism of the signal transmission and reception, formulates the ADR based on the mutual information and presents the analytic comparison with the channel capacity at low SNR.

\subsection{Communication Scheme}
\label{secTx}
Consider two independent BPSK-modulated symbol streams, $\textbf{x}_1 = [x_{1,1}, x_{1,2}, \cdots]$ and $\textbf{x}_2 = [x_{2,1}, x_{2,2}, \cdots]$, to be transmitted in parallel manner over a memoryless AWGN channel, where the BPSK symbols $x_{1,k}, x_{2,k} \in \{+1, -1\}$ for $k = 1, 2, \cdots$.

For simplicity, we shall omit the symbol's index, $k$, of BPSKs and work on the 4 possible combinations of the two symbols in categorization of two cases, wherein $x_1=-x_2$ belongs to case I and $x_1=x_2$ case II, of equal occurrence probability $\eta=1/2$.  

In each channel realization, the symbol $x_1$ is transmitted by adjusting its amplitude with respect to case I or II 
\begin{eqnarray}
\begin{array}{l}\label{eqx}
x = \left\{ \begin{array}{l}
\alpha {x_1}{\rm{\ \ \ \ \;if\ }}{x_1} = {x_2}{\rm{,\ case \ I,}}\\
\frac{1}{2}\beta {x_1}{\rm{\ \ \ if\ }}{x_1} \ne {x_2}{\rm{,\ case \ II,}}
\end{array} \right.
\end{array}
\end{eqnarray}
where $x$ is the transmitted signal, $\alpha$ and $\beta$ are two positive real numbers, to adjust the amplitude of $x_1$, with the assumption of $(\alpha > \beta >0)$. Transmitted signals are listed in the 4th column of Table {\ref{Table1}}. 

Under the assumption that $\beta$ has been known to the receiver, signal detection formulas are designed for detections of $x_1$ and $x_2$, separately, as expressed below.

The symbol $x_1$ is detected first by  
\begin{equation} \label{eqy1}
y_1 = x + n,
\end{equation}
then the symbol $x_2$ by 
\begin{equation} \label{eqy2}
y_2 = y_1 - \beta x_1 ,
\end{equation}
at the receiver, where $y_1$ is the received signal and $y_2$ is the modified signal of $y_1$ for detections of $x_1$ and $x_2$, $n$ is the AWGN component. Values of $y_1$ and $y_2$ are listed in the 5th- and 6th column of Table {\ref{Table1}}, respectively.   

\begin{table}[htb]
	\renewcommand{\arraystretch}{1.5}
	\centering
	\small
	\caption{The parallel transmissions of Cocktail BPSK symbols.}
	\label{Table1}
	\begin{tabular}{|c|c|c|c|c|c|}
		\hline
		\tabincell{c}{Case} & \tabincell{c}{$x_1$} & \tabincell{c}{$x_2$} & \tabincell{c}{$x$ }&\tabincell{c}{$y_1$}&\tabincell{c}{$y_2$} \\
		\hline		
		\multirow{2}*{I} & \tabincell{c}{$+1$} & \tabincell{c}{$+1$} & \tabincell{c}{$\alpha$}&\tabincell{c}{$\alpha+n_0$}&\tabincell{c}{$(\alpha-\beta)+n_0$} \\
		\cline{2-6}
		~ & \tabincell{c}{$-1$} & \tabincell{c}{$-1$} & \tabincell{c}{$-\alpha$}&\tabincell{c}{$-\alpha+n_0$} &\tabincell{c}{$-(\alpha-\beta)+n_0$}\\
		\hline		
		\multirow{2}*{II} & \tabincell{c}{$-1$} & \tabincell{c}{$+1$} & \tabincell{c}{$-\beta/2$}&\tabincell{c}{$-\beta/2+n_0$}&\tabincell{c}{$\beta/2+n_0$} \\
		\cline{2-6}
		~ & \tabincell{c}{$+1$} & \tabincell{c}{$-1$} & \tabincell{c}{$\beta/2$}&\tabincell{c}{$\beta/2+n_0$}&\tabincell{c}{$-\beta/2+n_0$}\\
		\hline
		
	\end{tabular}
\end{table}

In contrast to the conventional modulated signals, the symbol $x_2$ does not require any power at the transmitter, because it can reuse the symbol energy of $x_1$ at the receiver to its BPSK detection explained below.  

In the detection of $x_1$, (\ref{eqy1}) is used by $y_1>0$ or $y_1<0$ for making a decision of $x_1=+1$ or $x_1=-1$, respectively. 

Then, the result of $x_1$ is taken to (\ref{eqy2}) for the detection of $x_2$. Because that equation (\ref{eqy1}) can be an  error-free transmission as explained in the next subsection, the error propagation from $x_1$ can be precluded in the following discussions. 

The detection of $x_2$ can be carried out by $y_2>0$ or $y_2<0$ for making a decision of $x_2=+1$ or $ x_2=-1$, respectively. 

The energy reuse of the proposed method can be immediately found by viewing the input energy, which is exactly of the same value as that of the averaged symbol energy of $x_1$ with the two cases, i.e., case I and II, 
\begin{equation} 
\begin{array}{l} \label{eqEin}
E_{in} = E_1=\eta\alpha^2+ (1-\eta)( {\frac{1}{2}\beta})^2 ,
\end{array}
\end{equation}
where $\eta=1/2$ is the occurrence probability of case I, $E_{in}$ is the input energy and $E_1$ is the symbol energy of $x_1$. From the conventional point of view, the energy in (\ref{eqEin}) has been fully used in the transmission of $x_1$.

However, we have additionally symbol $x_2$ conveyed by the amplitudes of $(\alpha-\beta)$ and $\beta/2$ in case I and II, respectively. The reused energy can be found by 
\begin{equation} 
\begin{array}{l} \label{eqE2}
E_2 =\eta(\alpha-\beta)^2+ (1-\eta)( {\frac{1}{2}\beta})^2 ,
\end{array}
\end{equation}
where $E_2$ is the averaged symbol energy of $x_2$. 
  
In comparison with the convention methods, we can find an energy gain by 
\begin{equation} \label{eqdE}
\Delta E = E_T- E_{in} = E_2 ,
\end{equation}
with $E_T=E_1+E_2$, where $E_T$ is the total energy averaged over the two cases in signal detections of the proposed scheme and $\Delta E$ is defined as the energy gain factor in this paper. 
 
We refer the proposed method to as the cocktail BPSK, because the total energy of $x_1$ and $x_2$ can be attributed to the use of $\alpha$ and $\beta$, which adjust the energy ratio between the two BPSKs.  

Since the cocktail BPSK restricts its construction in one dimension, using its vertical dimension to set up another cocktail BPSK to increase the bandwidth efficiency is straightforward.   

\subsection{Theoretical Comparison}
The spectral efficiency analysis of the cocktail BPSK can start from (\ref{eqy1}), in which the detection of $x_1$ works on the two possible amplitudes of $x$, i.e., $\alpha$ and $\beta/2$, in case I and II. The ADR can be calculated by the mutual information 
\begin{eqnarray}
\begin{array}{l}\label{eqR1}
\mathbb{R}_1= \eta\mathbb{R}_\textrm{BPSK}(\alpha, \sigma_N^2) + (1-\eta) \mathbb{R}_\textrm{BPSK}(\frac{1}{2}\beta, \sigma_N^2),
\end{array}
\end{eqnarray} 
where $\mathbb{R}_1$ is the averaged ADR of $x_1$ and the function $\mathbb{R}_\textrm{BPSK}(\cdot, \cdot) $ is the mutual information of BPSK, which is explained in the appendix. 

It is noted that equation (\ref{eqR1}) describes the ADR of $x_1$ for the error-free transmission, which theoretically requires to work with an infinitive long data stream.  In this theoretical research, we use the assumption of error-free transmission to calculate the ADR of $x_2$ by taking the result of $x_1$ into (\ref{eqy2})  
\begin{eqnarray}
\begin{array}{l}\label{eqR2}
\mathbb{R}_2= \eta\mathbb{R}_\textrm{BPSK}(\alpha-\beta, \sigma_N^2) + (1-\eta) \mathbb{R}_\textrm{BPSK}(\frac{1}{2}\beta, \sigma_N^2),
\end{array}
\end{eqnarray}
where $\mathbb{R}_2$ is the averaged ADR of $x_2$.  

The spectral efficiency of this approach can be estimated by the summation of 
\begin{eqnarray}
\begin{array}{l}\label{eqR}
\mathbb{R}= \mathbb{R}_1+\mathbb{R}_2 ,
\end{array}
\end{eqnarray}
where $\mathbb{R}$ is the total ADR of the cocktail BPSK scheme.  

Using Taylor's expansion of (\ref{eqR}) to subtract (\ref{eqC2}) with the assumption of $\eta=1/2$ yields  
\begin{eqnarray}\label{eqdC}
\Delta C = ({\Delta E} / {\sigma _N^2}){\log _2}e ,
\end{eqnarray}
where $\Delta C$ is the ADR gain in comparison with both of the BPSK and the channel capacity at very low SNR.  

The authors note that the energy gain factor, $\Delta E$, places the key role for achieving the high spectral efficiency beyond the channel capacity.

\section{Numerical Results}

\begin{figure}[!t]
	\centering
	\subfigure[]{
		\includegraphics[width=0.5\textwidth]{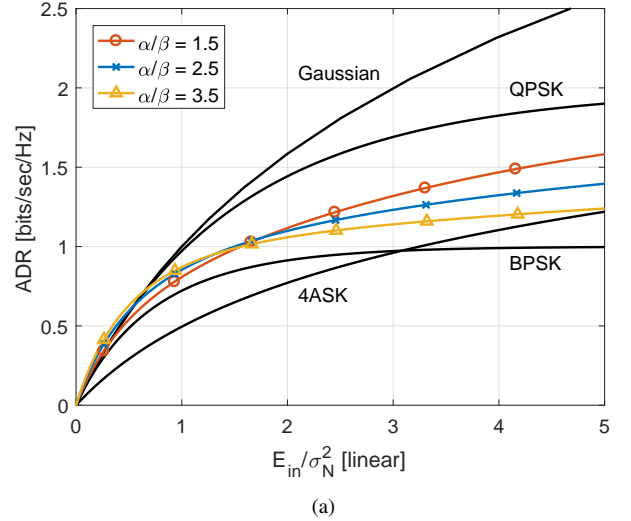}
		\label{fig2a}}
	\subfigure[]{
		\includegraphics[width=0.5\textwidth]{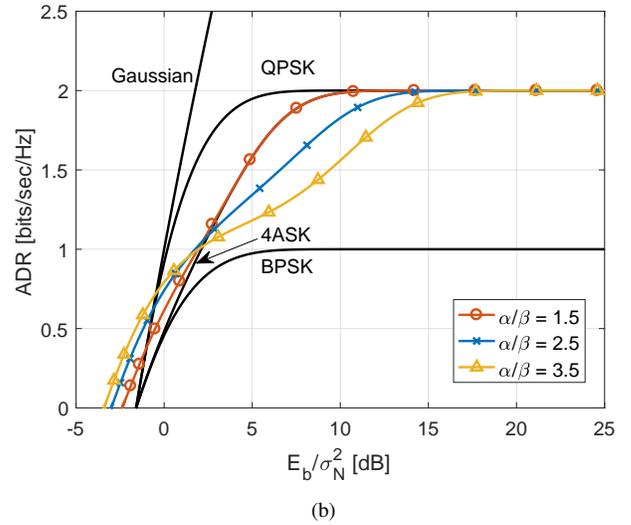}
		\label{fig2b}}
	\caption{Data rate comparisons between the proposed cocktail BPSK (C-BPSK) and conventional transmission schemes over AWGN channels versus (a) $E_{in}/\sigma_N^2$ [linear] and (b) $E_b/\sigma_N^2$ [dB].}
	\label{fig2}
\end{figure}

\begin{figure}[!t]
	\centering
	\subfigure[]{
		\includegraphics[width=0.45\textwidth]{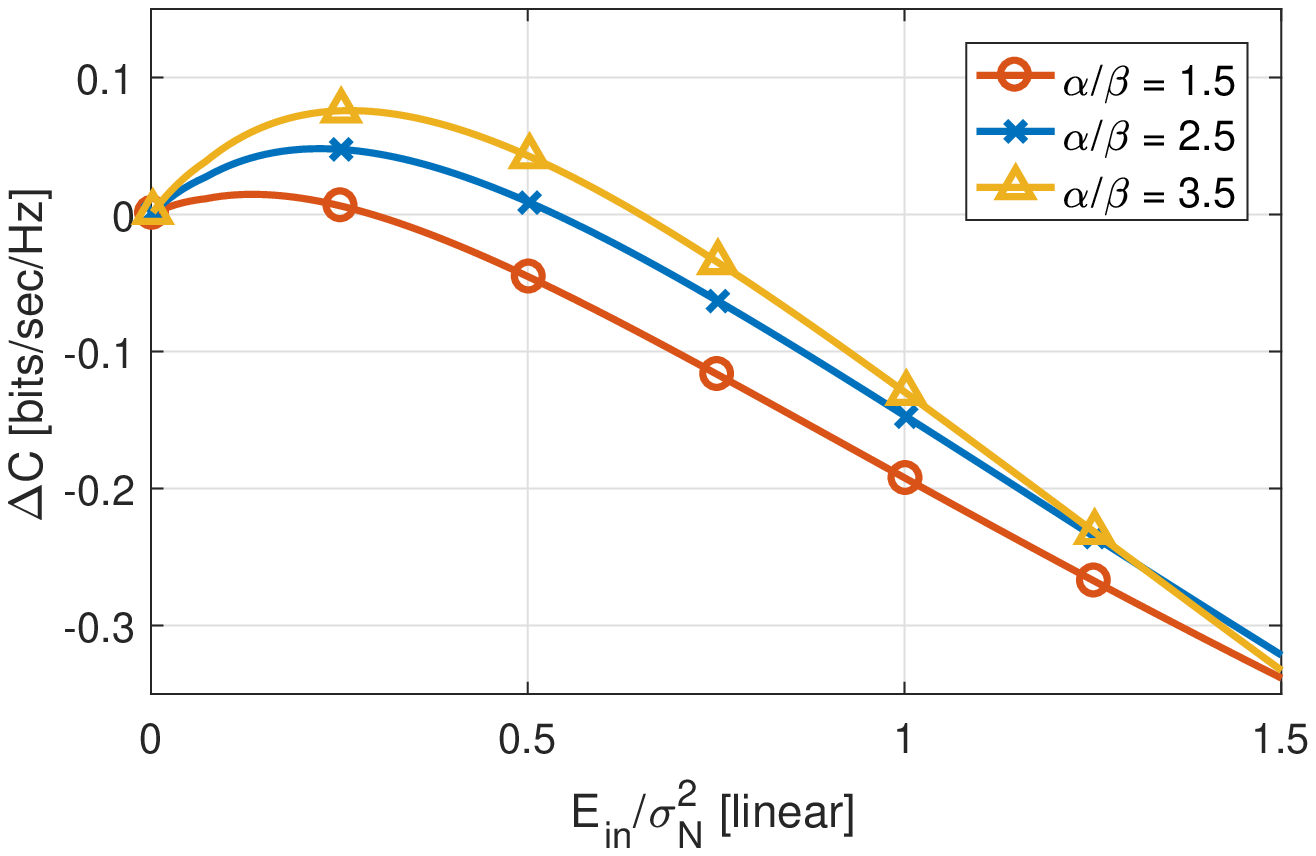}
		\label{fig3a}}
	\subfigure[]{
		\includegraphics[width=0.45\textwidth]{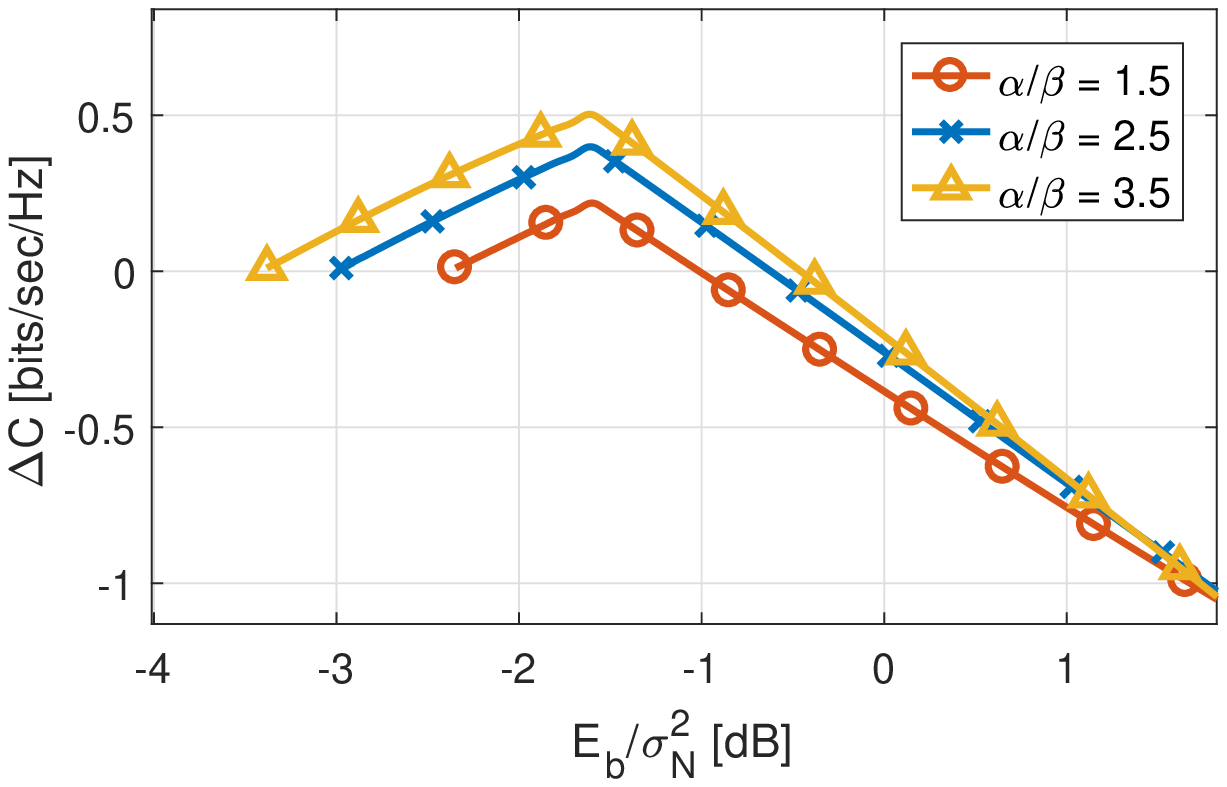}
		\label{fig3b}}
	\caption{Data rate differences between the proposal cocktail BPSK and Gaussian-distributed inputs versus (a) $E_{in}/\sigma_N^2$ [linear] and (b) $E_b/\sigma_N^2$ [dB].}
	\label{fig3}
\end{figure}

Herein, we compare the cocktail BPSK with the conventional BPSK for the ADR performances by setting the same input-energy/symbol-energy in the both methods.  

To express the ADR appropriately for the cocktail BPSK, we use the linear SNR, i.e., $E_{in}/{\sigma_N^2}$, at the horizontal axis. The numerical results of (\ref{eqR}) are taken to the comparison as shown in Fig. {\ref{fig2a}} for the various ratios of $\alpha$ to $\beta$.  

In addition, for better understanding, we convert the linear SNR in Fig. {\ref{fig2a}} to the conventional manner by {\cite{Es2Eb}}
\begin{eqnarray} \label{EbEs}
E_b/{\sigma_N^2} = \frac{E_{in}/{\sigma_N^2}}{\mathbb{R}} ,
\end{eqnarray}
and use decibel measurement at the horizontal axis, where $E_b$ is the bit-energy of the conventional BPSK and QPSK. The numerical results are plotted as shown in Fig. {\ref{fig2b}}.  

Viewing the ADRs curves horizontally in Fig. {\ref{fig2}} for a value at the vertical axis, we can find that the ADR curves of the cocktail BPSK can be on the left side of the channel-capacity-curve when the SNRs are very low. This indicates the outperforming of this approach compared to the channel capacity. The largest gain of 1.8 dB is found by the use of the ratio $\alpha$ to $\beta=3.5$ as shown in Fig. {\ref{fig2b}}, when the ADR is close to zero.

To show the results more distinctly, numerical results of (\ref{eqdC}) are plotted by two measurements, i.e., the linear SNR and the decibel measurement in Fig. {\ref{fig3a}} and (b), respectively. Spectral gains are found again when SNR is very low. However, as the SNR increases, values of the gain become smaller and, even, negative. The reason is explained below.


The ADR performance of the cocktail BPSK is dominated by that of BPSK's function (see the appendix), because it consists of 4 BPSKs with difference SNRs as the arguments in the function.  When the SNR increases, the ADRs of the all cocktail BPSKs go up.  In comparison with performances among cocktail BPSKs. we can find that the use of the larger ratio of $\alpha$ to $\beta$ performs better than that of the smaller ones at low SNR, while  worse at high SNR.  

The reason can be found by examining the SNR values in BPSK components of the cocktail BPSK: the component having the relative higher SNR argument can 
bring larger contribution to ADR at the very low SNR due to the linear property of the ADR as explained in (\ref{eqC2}), however, it can reduce its contribution faster when the SNR increases because of the concave-downward nature of BPSK.  Actually, the ADR performance derails from the linearity and becomes increasingly concave-downward, when SNR goes away from the zero.  The higher the SNR argument of the BPSK component is, the less contribution we can have at the high SNR.  

As can be found in (\ref{eqy1}) and (\ref{eqy2}), using a large ratio of $\alpha$ to $\beta$ can lead to a larger difference of amplitudes of BPSK components. Consequently, the SNR differences of BPSK components are also larger. This means that the cocktail BPSK of large value of $\alpha/\beta$ will contain the BPSK component of relative higher SNR arguments.   When SNR increases, these components show their disadvantages in the performance and the total up-going ADR will be more encumbered in comparison with those of smaller ratio of $\alpha$ to $\beta$. 

Finally, the ADR performances are saturated at the very high SNR due to the limitation from freedom degrees of modulated symbols.


\section{Conclusion}
In this paper, the cocktail BPSK was proposed by configuring two independent symbol streams at the transmitter on a non-orthogonal basis. In contrast to the conventional modulation schemes, this approach enables the energy reuse between the two symbols and gains energy at the receiver. Theoretical work proves that the proposed method can achieve higher ADR beyond the channel capacity at low SNR.  In addition, the cocktail BPSK is discussed for performances at both low- and high SNR. Numerical results of the mutual information confirm this approach.

\appendix

The analysis framework with the achievable data rate of conventional BPSK that is characterized by a function of the BPSK amplitude $A$ and the AWGN power $\sigma_N^2$, expressed as~\cite{Shannon1948}
\begin{equation}
\begin{array}{l}\label{rateBPSK}
\mathbb{R}_\textrm{BPSK}(A,{\sigma_N^2}) = {\rm{H}}(Y) - {\rm{H}}(N) \\
= - \int_{-\infty }^{ + \infty } {p(y){{\log }_2}p(y){\rm{d}}y} - {\log _2}(\sqrt {2\pi e{\sigma_N^2}} ) ,
\end{array}
\end{equation}
where ${\rm{H}}(Y) = \mathcal{E} \{- \log_2 p(y)\}$ is the entropy of the received signal with the probability density function given by
\begin{equation}
	p(y) = \frac{1}{2}\frac{1}{\sqrt{2\pi \sigma_N^2}}\left(e^{-\frac{(y-A)^2}{2 \sigma_N^2}} + e^{-\frac{(y + A)^2}{2 \sigma_N^2}}\right).
\end{equation}
Here, $\mathcal{E} \{\cdot \}$ denotes the expectation operator. Moreover, ${\rm{H}}(N) = {\log _2} (\sqrt{2 \pi e \sigma_N^2})$ is the entropy of the AWGN.


\begin{thebibliography}{1}

\bibitem{Shannon1948}
C. E. Shannon, ``A mathematical theory of communication'', \textit{The Bell System Technical Journal}, vol. 27, no. 3, pp. 379-423, July 1948.

\bibitem{EbN0}
S. Verdu, ``Spectral efficiency in the wideband regime'', \textit{IEEE Trans. Inform. Theory}, vol. 48, no. 6, pp. 1319-1343, June 2002.


\bibitem{UngerIT}
G. Ungerboeck, ``Channel coding with multilevel/phase signals'', \textit{IEEE Trans. Inform. Theory}, vol. IT-28, pp. 56-67, Jan. 1982.

\bibitem{Blahut}
R. E. Blahut, \textit{Principles and practice of information theory}. Boston: Addison-Wesley Longman Publishing Co., Inc., pp. 276-279, 1987.

\bibitem{math}
M. Abramowitz and I. A. Stegun (eds.), \textit{Handbook of mathematical functions with formulas, graphs, and mathematical tables}. New York: Dover Publications, Inc., 1972, pp.67-68.

\bibitem{jiao}
B. Jiao and D. Li, ``Double-space-cooperation method for increasing channel capacity'', \textit{China Communications}, vol. 12, no. 12, pp. 76-83, Dec. 2015.

\bibitem{ortho1}
A. J. Viterbi, \textit{CDMA: Principles of Spread Spectrum Communication}. Prentice Hall, 1995.

\bibitem{ortho2}
T. S. Rapporteur, \textit{Wireless Communications: Principles and Practice}. Prentice Hall, 2002.

\bibitem{Es2Eb}
J. M. Geist, "Capacity and cutoff rate for dense M-ary PSK constellations", \textit{Proc. Military Commun. Conf. (MILCOM)}, vol. 2, pp. 768-770, 1990-Sep. 30Oct. 3.


\end{thebibliography}
\end{document}